\documentclass[12pt,titlepage]{article}
\usepackage[onehalfspacing]{setspace}

\usepackage[margin=1in]{geometry}
\usepackage[utf8]{inputenc} % allow utf-8 input
\usepackage[T1]{fontenc}    % use 8-bit T1 fonts
\usepackage[english]{babel}
\usepackage{microtype}      % microtypography
\usepackage{hyperref}       % hyperlinks
\usepackage{url}            % simple URL typesetting
\usepackage{booktabs}       % professional-quality tables
\usepackage{amsfonts}       % blackboard math symbols
\usepackage{nicefrac}       % compact symbols for 1/2, etc.
\usepackage{lipsum}
\usepackage{authblk}
\usepackage{graphicx}
\usepackage{caption}
\usepackage{subcaption}
\usepackage{placeins}
\usepackage{soul}
\usepackage{url}
\usepackage{caption}
\usepackage{graphicx}
\usepackage{amsmath}
\usepackage{booktabs}
\usepackage{amsmath,amssymb}
\usepackage{mathtools}
\usepackage{nicefrac}
\usepackage[ruled,vlined]{algorithm2e}
\usepackage[numbers]{natbib}

\hypersetup{
    colorlinks=true,
    citecolor=blue,
    linkcolor=blue,
    urlcolor=blue
}
\title{Evolution of collective fairness in complex networks through degree-based role assignment}

\author[1,2]{Andreia Sofia Teixeira}
\author[1]{Francisco C. Santos}
\author[1]{Alexandre P. Francisco}
\author[3]{Fernando P. Santos}

\affil[1]{\small INESC-ID and Instituto Superior Técnico, Universidade de Lisboa, R. Alves Redol 9, 1000-029 Lisboa, Portugal}
\affil[2]{Indiana Network Science Institute, Indiana University, 1001 IN-45 Bloomington IN, USA}
\affil[3]{Department of Ecology and Evolutionary Biology, Princeton University, Princeton, NJ 08544, United States}

\begin{document}

\maketitle 

\begin{abstract}
From social contracts to climate agreements, individuals engage in groups that must collectively reach decisions with varying levels of equality and fairness. These dilemmas also pervade Distributed Artificial Intelligence, in domains such as automated negotiation, conflict resolution or resource allocation. As evidenced by the well-known Ultimatum Game --- where a Proposer has to divide a resource with a Responder --- payoff-maximizing outcomes are frequently at odds with fairness. Eliciting equality in populations of self-regarding agents requires judicious interventions. Here we use knowledge about agents’ social networks to implement fairness mechanisms, in the context of Multiplayer Ultimatum Games. We focus on network-based role assignment and show that preferentially attributing the role of Proposer to low-connected nodes increases the fairness levels in a population. We evaluate the effectiveness of low-degree Proposer assignment considering networks with different average connectivity, group sizes, and group voting rules when accepting proposals (e.g. majority or unanimity). We further show that low-degree Proposer assignment is efficient, not only optimizing fairness, but also the average payoff level in the population. Finally, we show that stricter voting rules (i.e., imposing an accepting consensus as requirement for collectives to accept a proposal) attenuates the unfairness that results from situations where high-degree nodes (hubs) are the natural candidates to play as Proposers. Our results suggest new routes to use role assignment and voting mechanisms to prevent unfair behaviors from spreading on complex networks.

\providecommand{\keywords}[1]
{
  \small	
  \textbf{\textit{Keywords---}} #1
}
\keywords{
Modeling and Simulation; Social networks; Evolutionary Game Theory; Population dynamics; Complex Systems; Fairness; Wealth inequality
}
\end{abstract}

%%%%%%%%%%%%%%%%%%%%%%%%%%%%%%%%%%%%%%%%%%%%%%%%%%%%%%%%%%%%%%%%%%%%%%%%

\section{Introduction}

Fairness has a profound impact on human decision-making and individuals often prefer fair outcomes over payoff maximizing ones~\cite{fehr2003nature}. This evidence is pointed through behavioral experiments, frequently employing to the celebrated Ultimatum Game (UG)~\cite{guth1982experimental}. In the UG, one Proposer decides how to divide a given resource with a Responder. The game only yields payoff to the participants if the Responder accepts the proposal. Human Proposers tend to sacrifice their share by offering high proposals and Responders often prefer to earn nothing rather than accepting unfair divisions. These counter-intuitive results motivated several theoretical models that aimed at justifying, mathematically, the evolution of fair intentions in human behavior~\cite{page2000spatial,nowak2000fairness,de2008learning,rand2013evolution}.%\\ 

In Distributed Artificial Intelligence and Multiagent Systems, fairness concerns are important in domains that go beyond pairwise interactions. Autonomous agents have to take part in group interactions that must decide upon outcomes possibly favoring different parts unequally. Examples of such domains are automated bargaining~\cite{jennings2001automated}, conflict resolution~\cite{pritchett2017negotiated} or multiplayer resource allocation~\cite{chevaleyre2006issues}. To capture some of the dilemmas associated with fairness versus payoff maximization in these interactions, we use a multiplayer extension of the Ultimatum Game~\cite{santos2015evolutionary} (MUG). Here, a proposal is made by a Proposer to a group of $N-1$ Responders that, collectively, decide to accept or reject it. As in the pairwise UG, the strategy of a Proposer, $p$, is the fraction of resource offered to the Responder; the strategy of the Responder, $q$, is the personal threshold used to decide about acceptance or rejection~\cite{nowak2000fairness,page2000spatial}. Groups decide to accept or reject a proposal through functions of the individual acceptance thresholds, $q$.  Group acceptance depends on a decision rule: if the fraction of acceptances equals or exceeds a minimum fraction of accepting Responders, $M$, the proposal is accepted by the group. In that case, the Proposer keeps what she did not offer ($1-p$) and the offer is divided by the Responders --- each receiving $p/(N-1)$. If the fraction of acceptances remains below $M$, the proposal is rejected by the group and no one earns anything. As in the UG, the \textit{sub-game perfect equilibrium} of MUG consists in a very low value of proposal $p$ and very low values of threshold $q$~\cite{santos2016dynamics}. %\\

Previous studies with the UG~\cite{page2000spatial,nowak2000fairness,de2008learning,rand2013evolution} and the MUG~\cite{santos2015evolutionary,santos2019evolution}, assume that the roles of Proposer and Responder are attributed following uniform probability distributions: each agent has the same probability of being selected to play as Proposer. These assumptions are naturally at odds with reality. In real-life Ultimatum Games, being the Proposer or the Responder depends on particular agents' characteristics. Proposers, such as employers, investors, auction first-movers or rich countries, are in the privileged position of having the material resources to decide upon which divisions to offer. This advantageous role is notorious if, again, one considers the theoretical prediction of payoff division in the UG (\textit{sub-game perfect equilibrium}) posing that Proposers will keep the largest share of the resource being divided. The benefits of Proposers are more evident when proposals are made to groups, as Responders need to divide the offers --- thus increasing the gap in gains between the single Proposer and the Responders. In this multiplayer context, punishing Proposers becomes harder: any attempt to punish unfair offers is only effective if there is a successful collective agreement --- on Responders --- to sacrifice individual gains and reject an offer. Asserting that these two roles are asymmetric, so should be the criteria to assign them, leading us to two main questions:

\begin{itemize}
\item How should a Proposer be selected within a group, in multiplayer ultimatum games to guarantee efficiency and fairness?
\item In networked games, which network-based criteria can be used to maximize long-term efficiency and fairness?%\\
\end{itemize}

Here we introduce a model, based on evolutionary game theory (EGT)~\cite{weibull1997evolutionary,tuyls2007evolutionary} and complex networks, to approach the previous questions. We analyze multiplayer ultimatum games in heterogeneous complex networks through network centrality-based role assignment. The fact that networks are heterogeneous allows us to test several node properties and centrality measures as base criteria for defining how to select Proposers in a group. We focus on degree centrality. We find that selecting low-degree Proposers elicits fairer offers and increases the overall fitness (average payoff) in a population. 

\section{Related Work}

The questions we address in this work --- and the model proposed to tackle them --- lie on the interface between mechanism for fairness elicitation in multiagent systems, multilayer bargaining interactions, dynamics on complex networks and network interventions to sustain socially desirable outcomes.%\\

Some of the most challenging contexts to elicit fairness involve the tradeoff between payoff-maximizing and fair outcomes. As stated, the UG~\cite{guth1982experimental} has been a fundamental interaction paradigm to study such dilemmas. In this context, reputations~\cite{nowak2000fairness} and stochastic effects~\cite{rand2013evolution} were pointed as mechanisms that justify fair behaviors.~\citeauthor{page2000spatial} found that, in a spatial setting, fairer proposals emerge as clusters of individuals proposing high offers are able to grow~\cite{page2000spatial}. Also in the realm of interaction networks,~\citeauthor{de2008learning} concluded that scale-free networks allow agents to achieve fairer agreements; rewiring links also enhances the agents’ ability to achieve fair outcomes~\cite{de2008learning}. A game similar to the UG assumes that Responders are unable to reject any proposal and Proposers unilaterally decide about a resource division. This leads to the so-called Dictator Game. In this context, reputations and mechanisms based on partner choice were also pointed as drivers of fair proposals~\cite{zisis2015generosity}. %\\

The previous works assume that all agents have the same probability of playing in the role of Proposer or Responder. Going from well-mixed (i.e., all individuals are free to interact with everyone else) to complex networks, however, provides the opportunity to implement network-based role assignment that considers network measures. In this context,~\citeauthor{wu2013adaptive} studied the pairwise UG in scale-free networks, with roles being attributed based on network degree. The authors show that attributing the role of Proposer to high-degree nodes leads to unfair scenarios~\cite{wu2013adaptive}. Likewise,~\citeauthor{deng2014impact} studied role assignment based on degree, concluding that the effect of degree-based role assignment depends on the mechanism of strategy update~\cite{deng2014impact}. When considering a pairwise comparison based on accumulated payoffs and social learning (as we do in the present work), the levels of contribution in the population increase if lower-degree individuals have a higher probability of being the Dictators. Both works consider the pairwise Ultimatum Game.%\\

In this work we use a multiplayer version of the UG (MUG) proposed in~\cite{santos2015evolutionary}. Other forms of multiplayer ultimatum games can be found in~\cite{fehr1999theory,grimm2015impact,takesue2017evolution}.~\citeauthor{santos2017structural} studied this game in the context of complex networks, showing that fairness is augmented whenever networks (where the game is played) allow agents to exert a sufficient level of influence over each other, by repeatedly participating in each others' interaction groups. The authors also find that stricter group decision rules allow for fairer strategies to evolve under MUG. Here we use networks to define group formation as suggested in the previously mentioned work (and originally in~\cite{santos2008social}). 

Departing from previous works that study degree-based role assignment in pairwise Ultimatum Games \cite{wu2013adaptive,deng2014impact}, we focus on a multiplayer game. As mentioned, this version highlights the asymmetries in the Proposer and Responder role: here, Proposers are likely to receive a even higher share of payoffs than each Responder -- as the latter need to divide accepted offers -- and punishing unfair Proposers depends on a group decision by the Responders -- that naturally may call for extra coordination mechanisms. Also, differently than in \cite{wu2013adaptive,deng2014impact}, here we show that, whenever highly connected nodes are the natural candidates to play in the role of Proposer, stricter voting rules (i.e., imposing an accepting consensus as requirement for collectives to accept a proposal) attenuates reduces the emergent level of inequality.%\\

Finally, the approach we follow in this work is akin to testing network interventions for social good. In this realm, we shall underline a recent work that employs EGT --- as we do in the present paper --- to study interventions to sustain cooperation in complex networks~\cite{lynch2018fostering}. The authors conclude that local interventions --- i.e., based on information about the neighborhood of the affected node --- outperform global ones. A similar conclusion is presented in~\cite{pinheiro2018local}. Several works study social dilemmas on top of complex networks and stress the conditions leading, in this context, to socially desirable outcomes~\cite{ranjbar2014evolution,raghunandan2012sustaining,salazar2011emerging,pinheiro2018local}. 

\section{Model and Methods}
\label{methods}
Here we detail the proposed evolutionary game theoretical model to evaluate the effect of degree-based role assignment on fairness under MUG. We start by providing details on the payoff calculation under MUG.

\subsection{Multiplayer Ultimatum Game}
\begin{figure}[t]
\centering
\includegraphics[width=0.6\linewidth]{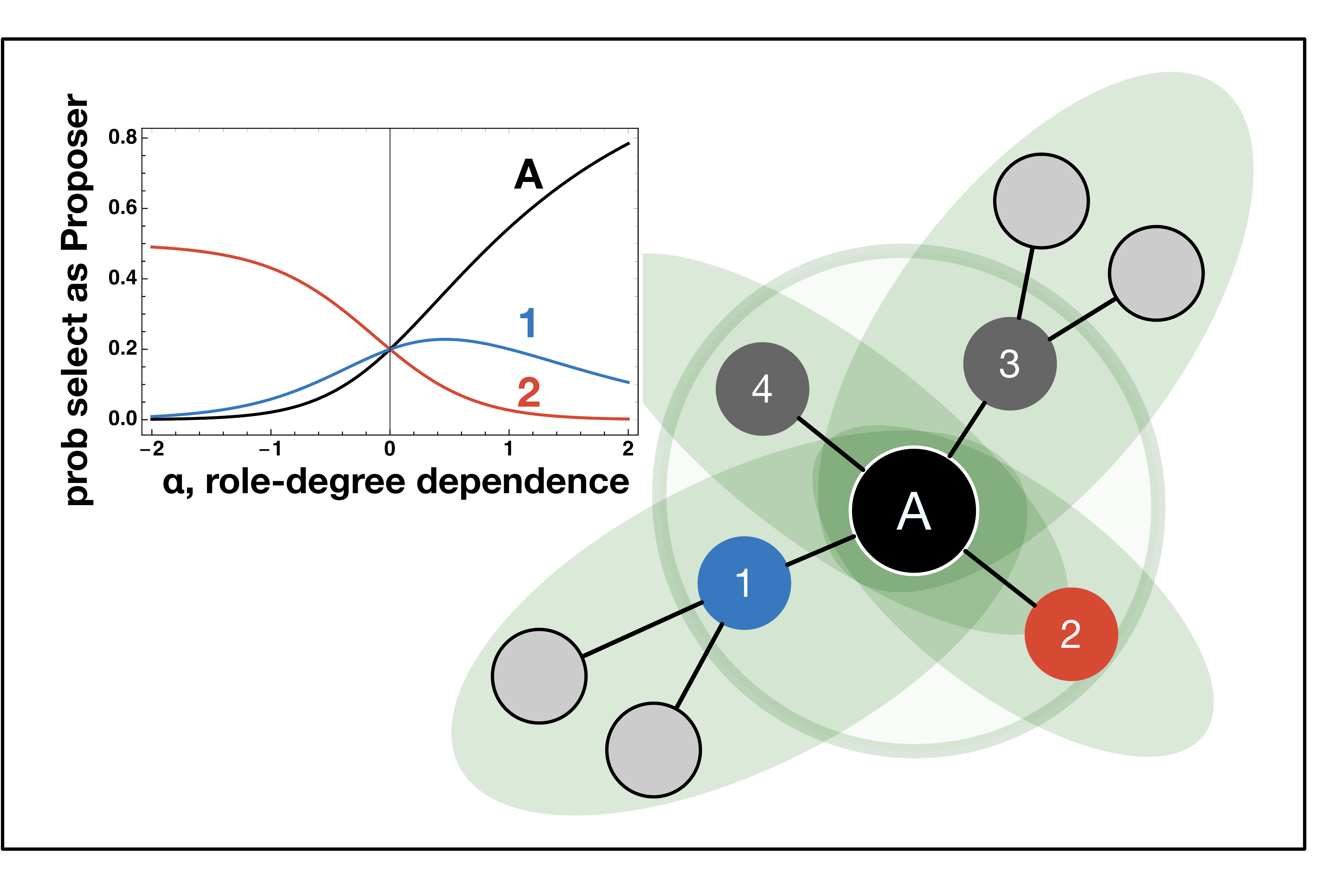}
\caption{Example of group formation and Proposer selection based on degree. Each node and its neighborhood define an interaction group. In the figure, node $A$ plays in 5 groups and its fitness results from the payoff sum after playing in all those groups. In general, a node plays in a number of groups equal to its degree plus one. For each group, the payoff is calculated after one individual is selected to be the Proposer. Proposer selection depends on the degree of each individual, and a parameter $\alpha$ controls this dependence (see Methods section). To exemplify this process, we represent the probability of each individual --- $A$ (high-degree), $1$ (medium-degree) and $2$ (low-degree) --- to be selected as a Proposer when playing in the group centered in $A$, as a function of $\alpha$.} 
\label{fig:1}
\end{figure}
In the 2-player UG, a Proposer has a resource and is required to propose a division with a Responder. The game only yields payoff to the participants if the Responder accepts the proposal~\cite{guth1982experimental}. Given a Proposer with strategy $p\in[0,1]$ and a Responder with strategy $q\in[0,1]$, the payoff for the Proposer yields

\begin{equation}
\Pi_P(p,q)=
\begin{cases}
1-p, p \geq q\\
0, p < q
\end{cases} ,
\end{equation}

and for the Responder

\begin{equation}
\Pi_R(p,q)=
\begin{cases}
p, p \geq q\\
0, p < q
\end{cases} .
\end{equation}

In the MUG, proposals are made by one Proposer to the remaining $N-1$ Responders, who must individually reject or accept it~\cite{santos2015evolutionary,santos2019evolution}. Since individuals may act both as Proposers and Responders (with a probability that will depend on node characteristics), we assume that each individual adopts a strategy ($p$, $q$). When playing as Proposer, individuals offer $p$ to the Responders. Responders will individually accept or reject the offer having their $q$ as a baseline: if the share of an offer $p$ is equal or larger than $q$ (i.e., $\frac{p}{N-1} \geq q$) the individual accepts the proposal. Otherwise, the Responder rejects that proposal. We can regard $q$ as the minimum fraction that an individual is willing to accept, relatively to the maximum to be earned as a Responder in a group of a certain size. Alternatively, we could assume that individuals ignore the group size and as such, when faced with a proposal, they must judge the absolute value of that proposal (an interpretation that also holds if we assume that individuals care about the whole group payoff). %\\

Overall group acceptance will depend upon $M$, the minimum fraction of Responders that must accept the offer before it is valid. Consequently, if the fraction of individual acceptances stands below $M$, the offer will be rejected. Otherwise, the offer will be accepted. In this case, the Proposer will keep $1-p$ to himself and the group will share the remainder, that is, each Responder gets $p/(N-1)$. If the proposal is rejected, no one earns anything. All together, in a group with size $N$ composed of $1$ Proposer with strategy $p \in [0,1]$ and $N-1$ Responders with strategies $(q_1, ..., q_{N-1}) \in [0,1]^{N-1}$ the payoff of the Proposer is given by 

\begin{equation}
\Pi_P(p,q_1,...,q_{N-1})=
\begin{cases}
1-p, \sum_{i=1}^{N-1}{\Theta(\frac{p}{N-1}-q_i)}/(N-1) \geq M\\
0, \text{otherwise}
\end{cases} ,
\end{equation}

where $\Theta(x)$ is the Heaviside step function, that evaluates to $1$ when $x\geq0$ and evaluates to $0$ when $x<0$. The payoff of any Responder in the group yields,

\begin{equation}
\Pi_R(p,q_1,...,q_{N-1})=
\begin{cases}
\frac{p}{N-1}, \sum_{i=1}^{N-1}{\Theta(\frac{p}{N-1}-q_i)}/(N-1) \geq M\\
0, \text{otherwise}
\end{cases}.
\end{equation}

We assume that MUG is played on a complex network, in which individuals are assigned nodes and links define who can interact with whom. Following~\cite{santos2008social,santos2017structural}, every neighborhood characterizes a N-person game, such that the individual fitness (or success) of an individual is determined by the payoffs resulting from the game centered on herself plus the games centered on her direct neighbors. We provide a visual representation of such group formation in Figure \ref{fig:1}. Degree heterogeneity will create several forms of diversity, as individuals face a different number of collective dilemmas depending on their degree (and social position); groups where games are played may also have different sizes. Such diversity is  introduced by considering two types scale-free networks. One is generated with the Barabási-Albert algorithm (BA) of growth and preferential attachment~\cite{barabasi1999emergence} leading to a power-law degree distribution, and a low clustering coefficient. The clustering coefficient offers a measure of the likelihood of finding triangular motifs or, in a social setting, how likely two friends of a given node are also friends of each other, a topological property of  relevance in the context of fairness and N-person games~\cite{santos2017structural}. In the second case, we consider the Dorogotsev-Mendes-Samukhin (DMS) duplication model~\cite{Dorogotsev2001}, exhibiting the same power-law degree distributions, yet with large values of the clustering coefficient.

\subsection{Networks generated}
In the BA model~\cite{barabasi1999emergence}, at each time step, the network grows by adding a new node and connecting it to \textit{m} other nodes already in the network. These connections are probabilistic, depending on the degree of the nodes to be connected with: having a higher degree increases the probability of having new connection. This process result in heterogeneous degree distributions, in which older nodes become highly connected (creating so-called \textit{hubs}). This is the combination of two processes -- \textit{growth} and \textit{preferential attachment}. In the DMS model~\cite{Dorogotsev2001}, at each time step, a node is added; instead of choosing other nodes to connect with, it chooses one edge randomly and connects to both ends of the edge. The networks generated by the DMS model have higher cluster coefficient than those with BA model, combining the high-clustering and high-heterogeneity that characterizes real-world social networks.

\subsection{Network based role selection}

Previous works show that anchoring the probability of nodes being selected for the role of Proposer or Responder on their degree has a sizable and non-trivial effect on the evolving levels of proposal in traditional two-person Ultimatum Games~\cite{wu2013adaptive,deng2014impact}. Considering multiplayer ultimatum games, however, opens space to study the interplay between group characteristics (such as group sizes) and network-based criteria to select Proposers in completely unexplored directions. So far, we assume that nodes are selected to be Proposers based on their degree. As such, in a group with $N$ individuals, where each individual $i$ has degree $k_i$, the probability that $j$ is selected as Proposer is given by $p_j=\frac{e^{\alpha k_j}}{ \sum_{i}{e^{\alpha k_i}}}$, where $\alpha$ controls the dependence of degree on role selection. One node is selected as Proposer and the remaining $N-1$ play as Responders.

\subsection{Evolutionary Dynamics}

We simulate the evolution of $p$ and $q$ in a population of size $Z$, much larger than the group size $N$. Initially, each individual has values of $p$ and $q$ drawn from a discretized uniform probability distribution in the interval $[0,1]$. The fitness $f_i$ of an individual $i$ of degree $k$ is determined by the payoffs resulting from the game instances occurring in $k+1$ groups: one centered on her neighborhood plus $k$ others centered on each of her $k$ neighbors (see~Figure \ref{fig:1}). Values of $p$ and $q$ evolve as individuals tend to imitate (i.e., copy $p$ and $q$) the neighbors that obtain higher fitness values. %\\

\begin{algorithm}[t]
\DontPrintSemicolon
\BlankLine
\small
    Initialize all $p_i, q_i \in Z = X \sim \mathcal{U}(0,1)$ \\
	\For(Main cycle of interaction and strategy update:){$t \leftarrow 1$ \KwTo $Gens$}{
		
		\For(Select agent to update:){$j \leftarrow 1$ \KwTo $Z$}{ 
			    \tcc{Sample two neighbors the population}
			    \emph{$A \leftarrow X \sim \mathcal{U}(1,Z_P)$} (agent to update)\\
			    \emph{$B \leftarrow Y \in neighbours(A)$} \\
			\If(Mutation:){$X \sim \mathcal{U}(0,1) < \mu$}{
				$p_A \leftarrow X \sim \mathcal{U}(0,1)$\\
				$q_A \leftarrow X \sim \mathcal{U}(0,1)$
			}
			
			\Else(Imitation:){
				%\emph{$B \leftarrow X \sim \mathcal{U}(1,Z)$, $B \neq A$}\\
				\emph{$f_A \leftarrow$ fitness($A$)}\\
				\emph{$f_B \leftarrow$ fitness($B$)}\\
				\emph{$prob \leftarrow \nicefrac{1}{\left(1+e^{-\beta(f_B-f_A)}\right)}$}\\
				\If{$X \sim \mathcal{U}(0,1) < prob$}{$p_A \leftarrow p_B$ + imitation error $\sim \mathcal{U}(-\varepsilon,\varepsilon)$ \\
				$q_A \leftarrow q_B$ + imitation error $\sim \mathcal{U}(-\varepsilon,\varepsilon)$}
			}
		}
	}
\caption{Pseudo-code of the main cycle of our simulations. We perform 100 runs over 10 different networks of each type (BA and DMS) with $2 \times 10^5$ generations per run.}\label{alg:syn}
\end{algorithm}\DecMargin{1em}

The numerical results presented below were obtained for structured populations of size $Z=1000$. {Similar results were obtained for $Z=10000$.} As already mentioned, we consider networks generated with both BA  and DMS algorithms, with average degree $\langle k\rangle = \{4,8,16\}$. Simulations take place for $2 \times 10^5$ generations, considering that, in each generation, all the individuals have (on average) the opportunity to revise their strategy through imitation. At every (discrete and asynchronous) time step, two individuals $A$ and $B$ (neighbors) are selected from the population. Given the group setting of the MUG, $B$ is chosen from one of the neighbours of $A$. Their individual fitness is computed as the accumulated payoff in all possible groups for each one, provided by the underlying structure (in each group the role of Proposer or Responder is selected following the section below); subsequently, $A$ copies the strategy of $B$ with a probability $\chi$ that is a monotonic increasing function of the fitness difference $f_B-f_A$, following the pairwise comparison update rule: $\chi=\frac{1}{1+e^{-\beta(f_B-f_A)}}$ \cite{traulsen2006stochastic}. The parameter $\beta$ specifies the selection pressure ($\beta=0$ represents neutral drift and $\beta$ represents a purely deterministic imitation dynamics). Imitation is myopic: the value of $p$ and $q$ copied will suffer a perturbation due to errors in perception, such that the new parameters will be given by $p'= p + \zeta_{p,\varepsilon}$ and $q' = q + \zeta_{q,\varepsilon}$, where $\zeta_{p,\varepsilon}$ and $\zeta_{q,\varepsilon}$ are uniformly distributed random variables drawn from the interval $[-\varepsilon,\varepsilon]$. This feature not only i) models a slight blur in perception but also ii) helps to avoid the random extinction of strategies, and iii) ensures a complete exploration of the strategy spectrum. To guarantee that new $p$ and $q$ are not lower than 0 or higher than 1, we implement reflecting boundaries at 0 and 1. Furthermore, with probability $\mu$, imitation will not occur and the individual will adopt random values of $p$ and $q$, proceeding through a random exploration of behaviors. We use $\mu=1/Z$, $\beta=10$ and $\varepsilon = 0.05$ throughout this work. The effect of varying $\mu$ is similar to the one verified when changing $\varepsilon$: an overall increase of randomness leads to higher chances of fairer offers (as in \cite{rand2013evolution,santos2015evolutionary}). For each combination of parameters, the simulations are repeated $100$ times (using $10$ different networks from each class studied), whereas each simulation starts from a population where individuals are assigned random values of $p$ and $q$ drawn uniformly from $[0,1]$. We provide a summary of the algorithm used to revise agents' strategies in Algorithm \ref{alg:syn}. The average values of $p$, $q$ and $f$ (denoted by $\langle p \rangle$, $\langle q \rangle$ and $\langle f \rangle$) are obtained as a time and ensemble average, taken over all the runs (considering the last $10^5$ generations, disregarding an initial transient period).

\section{Results}
\begin{figure}[h!]
\centering
\includegraphics[width=0.6\linewidth]{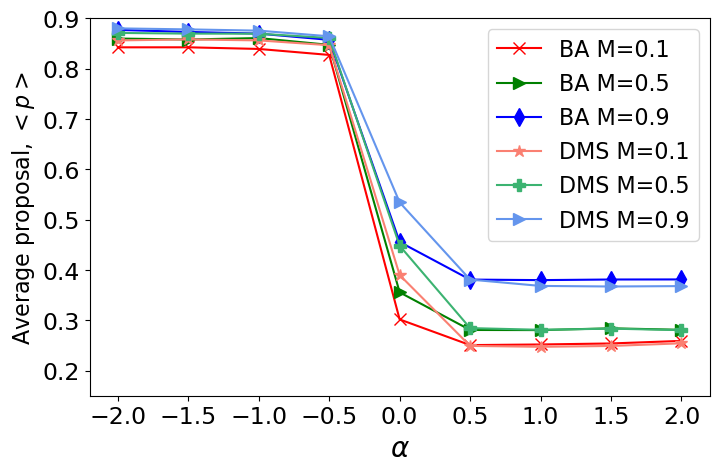}
\caption{The average proposal played by agents in a population, $\langle p \rangle$, decreases with $\alpha$. This means that attributing the role of Proposer to high-degree nodes reduces the overall fairness level in a population. We present results for BA and DMS networks with average degree $\langle k \rangle = 4$. We verify that low-degree Proposer assignment maximizes $\langle p \rangle$ for different group decision rules, $M = \{ 0.1, 0.5, 0.9\}$, i.e., the fraction of Responders that needs to accept a proposal for it to be accepted by the group.}
%We test two classes of heterogeneous networks: BA and DMS (see Methods for more details) with average degree $\langle z \rangle = 4$. We verify that low-degree Proposer assignment maximizes $\langle p \rangle$ for different group decision rules, $M = \{ 0.1, 0.5, 0.9\}$, i.e., the fraction of Responders that needs to accept a proposal for it to be accepted by the group.} 
\label{fig:2}
\end{figure}
\begin{figure*}[h!]
\centering
\includegraphics[width=1.0\linewidth]{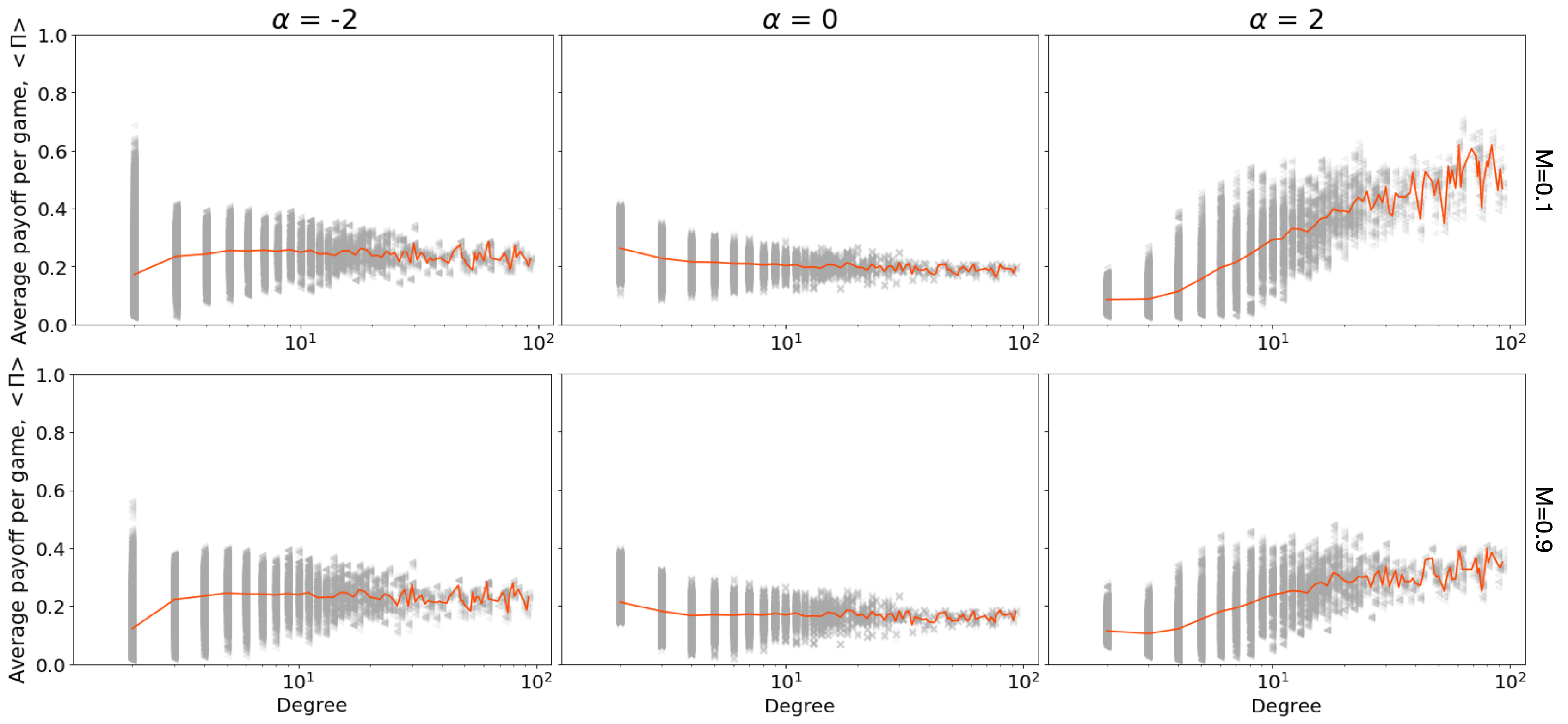}
\caption{On top of decreasing the average level of proposal in the population, $\langle p\rangle$, we found that attributing the role of Proposer to highly connected nodes decreases the level of fairness and equality \textit{within} the population. Here we use scatter plots to observe the average payoff obtained per game, $\langle \Pi \rangle$, for individuals with a certain degree (horizontal axis). The left panels represent a low-degree Proposer assignment scenario ($\alpha=-2$); the center panels represent random --- and degree-independent ---  role attribution ($\alpha=0$); the right panels represent a high-degree Proposer assignment scenario ($\alpha=-2$). Each gray cross represents a node in a degree-$\langle \Pi \rangle$ space; the orange line represents the mean taken over all nodes with a certain degree. Top panels represent $M=0.1$ and bottom panels $M=0.9$. High $\alpha$ --- i.e., high-degree Proposer assignment --- implies that highly connected nodes earn (approximately) five times more payoff per game than low-connected nodes (top-right panel). This effect is alleviated for higher $M$; for $M=0.9$, highly connected nodes earn (approximately) three times more payoff per game than low-connected nodes (bottom-right panel).} 
\label{fig:3}
\end{figure*}

We run the proposed model and record the average strategies played by the agents over time and over different runs (starting from different initial conditions, see Methods). We find that attributing the role of Proposer to low-degree nodes (or \textit{low-degree Proposer assignment}) increases the average level of proposal, $p$, adopted in the population of adaptive agents. This means that the payoff gap between Proposers and Responders is alleviated. Figure~\ref{fig:2} shows that, for low $\alpha$ ($\alpha < 0$), we obtain higher levels of average proposal when considering BA (low clustering coefficient) and DMS (high clustering coefficient) networks. We observe a steep decline in average proposals when the role of Proposer and Responder is attributed regardless the degree of individuals ($\alpha=0$). The low-proposal tendency is maintained if the role of Proposer is assigned to high-degree nodes ($\alpha>0$).%\\

%After running the proposed model and recording the average strategies played by agents over time and over different runs (starting from different initial conditions, see Methods) we found that attributing the role of Proposer to low-degree nodes (or \textit{low-degree Proposer assignment}) increases the average level of proposal, $p$, adopted in the population of adaptive agents. This means that the payoff gap between Proposers and Responders is alleviated. Figure~\ref{fig:2} shows that, for low $\alpha$ ($\alpha < 0$), we obtain higher levels of average proposal when considering BA (low clustering coefficient) and DMS (high clustering coefficient) networks. We observe a steep decline in average proposals when the role of Proposer and Responder is attributed regardless the degree of individuals ($\alpha=0$). The low-proposal tendency is maintained if the role of Proposer is assigned to high-degree nodes ($\alpha>0$).

We also confirm that high-degree Proposer assignment leads to unequal (unfair) results within a population. Figure~\ref{fig:3} depicts the average payoff gains for individuals with a certain degree. We can observe that, for $\alpha=2$, high-degree nodes obtain much higher values of payoff than low-degree nodes. This situation is ameliorated if individuals with lower degree are given a higher chance of becoming Proposers (lower $\alpha$) and, to a lower extent, if more Responders are required to accept a proposal in order for it to be accepted (higher $M$, bottom panels in Figure~\ref{fig:3}).%\\
%The previous results refer to the aggregated level of proposal in a population. We can also confirm that high-degree Proposer assignment leads to unequal (unfair) results within a population. Figure \ref{fig:3} depicts the average payoff gains for individuals with a certain degree. We can observe that, for $\alpha=2$ (right panels), high-degree nodes obtain much higher values of payoff than low-degree nodes. This situation is ameliorated if individuals with lower degree are given a higher change of becoming Proposers (lower $\alpha$) and, to a lower extent, if more Responders are required to accept a proposal in order for it to be accepted ($higher M$, bottom panels in Figure \ref{fig:3}).

\begin{figure*}[h!]
\centering
\includegraphics[width=\linewidth]{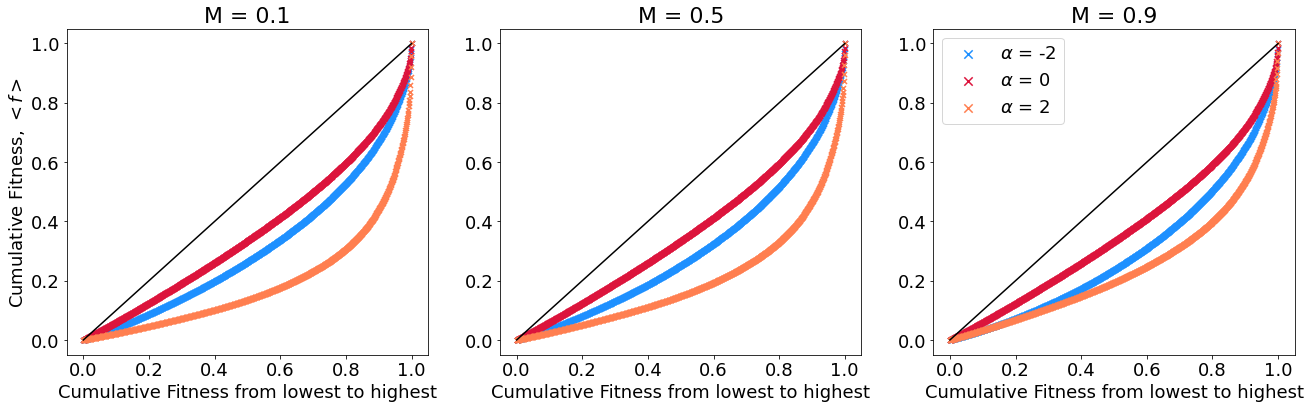}
\caption{Selecting high-degree nodes as Proposers increases unfairness. Here we represent the so-called Lorenz curves, often used to compute the Gini coefficients -- a typical measure of income inequality. Each curve is generated by ordering individuals by increasing value of income and plotting the corresponding cumulative distribution. Curves  closer to the perfect equality line (45 degree line) represent more egalitarian outcomes. Here we observe, yet again, that assigning the role of Proposer to high-connected nodes ($alpha=2$) yields unfair outcomes (orange line). While this is evident for soft (panel a, $M=0.1$), medium (panel b, $M=0.5$) and strict decision rules (panel c, $M=0.9$), we also verify that whenever hubs are the Proposers ($\alpha=2$), having strict decision rules (high $M$) reduces unfairness. %Other parameters ... 
} 
\label{fig:6}
\end{figure*}
\begin{figure*}[h!]
\centering
\includegraphics[width=1.0\linewidth]{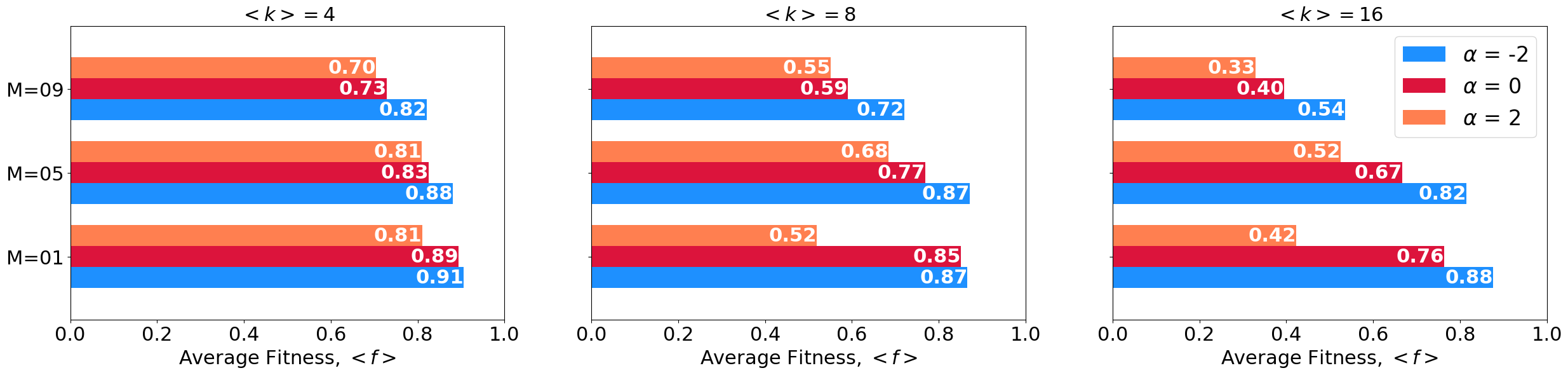}
\caption{Low-degree Proposer assignment maximizes the average fitness (i.e., sum of payoffs taken over all games, see Figure~\ref{fig:1}) in a population. Here we observe that the average fitness, $\langle f \rangle$, increases as $\alpha$ decreases. We show results for BA networks with different values of $\langle k\rangle$ and $M$. A similar conclusion is obtained when considering DMS networks with the same parameters as Figure~\ref{fig:2} and~\ref{fig:4}. %{With increasing $\langle k\rangle$, we observe a decline in the average fitness, as larger groups imply that the total payoff to be divided per game decreases. Nonetheless, more games are played attenuating this effect.}}
{Note that, increasing $\langle k\rangle$ implies that the average group size to play MUG also increases, which leads offers to be divided by larger groups (hence contributing to lower values of average payoff per game). On the other hand, increasing $\langle k\rangle$ means that more games are played, thus contributing to an increase in accumulated fitness (taken as the sum of payoffs in all games played).}}
\label{fig:4}
\end{figure*}

We can further verify the effect of $\alpha$ on fairness through the so-called \textit{Lorenz curves}~\cite{lorenz1905methods}, often used to compute the \textit{Gini coefficients}~\cite{gini1921measurement} that quantify income inequality. In Figure \ref{fig:6} we represent the Lorenz curves associated with different role-assignment rules ($\alpha$) and voting rules, $M$. Each curve is generated by ordering individuals by increasing value of income plotting the corresponding cumulative distribution. A curve closer to the perfect equality line (45 degree) represents a most egalitarian distribution of resources and a lower Gini coefficient. As we verify in Figure \ref{fig:6}, the most unequal outcomes (higher Gini) are obtained for higher $\alpha$. We further verify that, when fixing $\alpha=2$, having stricter voting rules (high $M$, in this case $M=0.9$) attenuates the unfairness associated with having hubs being the Proposers.%\\

%of $\langle k \rangle=4$ (left), $\langle k \rangle=8$ (center) and $\langle k \rangle=16$ (right). 
Not only does low-degree Proposer assignment reduce unfairness, as it also sustains more efficient outcomes --- taken as higher values of average fitness observed in the population. As stated previously, fairness represents here the payoff that a node obtains after playing in each possible group (see Figure~\ref{fig:1}). In Figure~\ref{fig:4} we confirm that low values of $\alpha$ maximize the average fitness of populations. This occurs when considering heterogeneous networks with different average degrees ($\langle k \rangle$) and group decision rules ($M$). This effect is more evident when considering less strict group decision rules (that is, lower $M$, meaning that less Responders are required to accept a proposal for the group to accept it) and networks with higher $\langle k \rangle$. %\\

\begin{figure}[t!]
\centering
\includegraphics[width=1\linewidth]{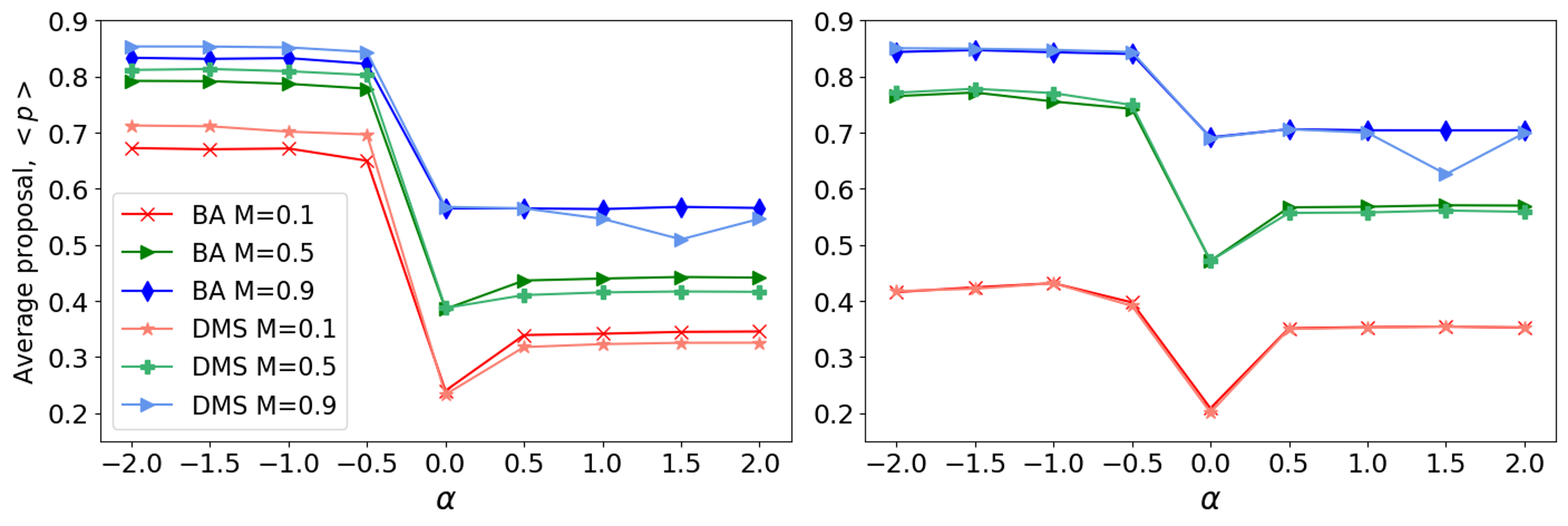}
\caption{We confirm that low-degree Proposer assignment maximizes populations' average level of proposal, $\langle p \rangle$, for both BA and DMS networks with higher average degree ($\langle k \rangle = 8$, top, and $\langle k \rangle = 16$, bottom). For networks with higher $\langle k \rangle$ --- leading to MUGs played in larger groups --- and low $M$, random role attribution ($\alpha=0$) configures the worst scenario in terms of fair proposals.} 
\label{fig:5}
\end{figure}
%This said, there is a non-linear effect between $\alpha$ and $M$: for low $M$, a degree-independent role assignment ($\alpha=0$) minimizes the observed proposal value in the population.

Finally, we confirm that low-degree Proposer assignment maximizes the average proposal played in the population (and thus fairness) when considering networks with higher $\langle k \rangle$ and, as a result, larger average group sizes. As Figure~\ref{fig:5} conveys, the higher values of average proposal, $\langle p \rangle$ are obtained for $\alpha<0$. Notwithstanding, we are able to find parameter spaces where the dependence of $\langle p \rangle$ on $\alpha$ is seemingly affected by \textit{i}) the average connectivity of the network --- and thus on the average size of the groups in which MUG is played --- and \textit{ii})  particular values of $M$. Also, we confirm that increasing $M$ increases $\langle p \rangle$ for all values of $\alpha$. Our results suggest that offering the first move to low-degree nodes balances the natural power of highly connected nodes in scale-free networks, leading to a significant increase in the global levels of fairness. Interestingly, we also find that particular voting rules ($M$) are able to attenuate the negative effect of high $\alpha$ (\textit{i.e.} privileged high-degree nodes being selected to be Proposers) on fairness.

\section{Discussion and Conclusion}

In this paper we address the general problem of 1) deciding how to attribute bargaining roles in a social network and, in particular, 2) understanding the impact of different criteria on the emerging levels of fairness in Multiplayer Ultimatum Games. We verified that attributing the role of proposer to low degree nodes boost both fairness and overall fitness. This conclusion remains valid for different network structures (BA and DMS networks with average degrees ranging from 4 to 16) and interaction scenarios (in terms of group sizes and group decision rules) . %\\

One possible intuition for this result can be understood as follows: consider the situation where two hubs ($H_1$ and $H_2$) are connected to each other and each is linked to $k-1$ nodes with degree $1$ (i.e., leafs); given that imitation is based on the accumulated payoff, these hubs have, potentially, a higher fitness than the leafs. When highly connected nodes are selected to be the Proposers, $H_1$ and $H_2$ always play in that role. If this is the case, the hubs always use the values of $p$ they adopt and the only way that leafs earn some payoff is to lower their acceptance thresholds and always accept hubs' offers. When low connected nodes are selected to be the Proposers, however, the hubs will always play in the role of Responder. Assuming that all leafs adopt strategy $p$, the fitness of $H_1$ and $H_2$ will be given by $2\frac{p}{k} + p(k-1)$ --- the first term corresponding to the payoff earned when playing in a group centered in $H_1$ or $H_2$ and the second term corresponds to the payoff earned when playing with each of the Proposer-leafs. Notice that the fitness of the hubs (and hence the probability of getting imitated) increases with $p$, the offers made by the leafs. As leafs are only connected with $H_1$ or $H_2$, the only way they have to spread their strategies --- which occurs when the hubs imitate each other --- is by increasing their offered values, $p$. Fairer strategies thus spread under low-degree Proposer assignment. %\\

We also find that the perils of having high-degree Proposers can be softened with strict group decision rules. This means that, whenever role selection is constrained and Proposers are necessarily the better connected nodes (by having the needed resources to propose and be the first movers in a bargaining situation) unfairness can be reduced by imposing that proposals need to be validated by a large fraction of Responders. The effect of $M$ on eliciting fairer offers is similar to that found in recent literature~\cite{santos2015evolutionary,santos2019evolution}. Also, our results are in line with works showing that selecting low-degree Proposers maximizes fairness in the context of pairwise Ultimatum Games~\cite{wu2013adaptive} and Dictator Games~\cite{deng2014impact}. %\\

This work can underlie several extensions of interest for social and engineering sciences. Here we consider that role assignment is endogenously imposed. In reality, these rules are likely to evolve side-by-side with individual strategies, being another self-organized property of the system, as fairness and wealth distributions. Moreover, the fact that network-based role assignment elicits fairness in rather complicated scenarios --- as multiplayer bargaining games --- suggests that such approach could also be used within the broader context of active interventions aiming at fostering fairness in hybrid populations comprising humans and machines~\cite{santos2019evolution,shirado2017locally,rahwan2019machine}. In this context, it would be relevant to assess --- both experimentally and through numerical simulations --- the impact on human decision-making of having virtual regulators dynamically deciding the role to adopt by their group peers, depending on their position in the interaction structure. Despite these open questions, our present work already suggests that carefully selecting the role of each agent within a group --- depending on their social position and without limiting their available options ---  can offer a long-term social benefit, both in terms of the overall levels of fairness, wealth inequality, and global wealth of a population comprised by self-regarding agents. %\\

\bibliographystyle{plainnat} 
\bibliography{sample}

\begin{thebibliography}{34}
\providecommand{\natexlab}[1]{#1}
\providecommand{\url}[1]{\texttt{#1}}
\expandafter\ifx\csname urlstyle\endcsname\relax
  \providecommand{\doi}[1]{doi: #1}\else
  \providecommand{\doi}{doi: \begingroup \urlstyle{rm}\Url}\fi

\bibitem[Barab{\'a}si and Albert(1999)]{barabasi1999emergence}
Albert-L{\'a}szl{\'o} Barab{\'a}si and R{\'e}ka Albert.
\newblock Emergence of scaling in random networks.
\newblock \emph{Science}, 286\penalty0 (5439):\penalty0 509--512, 1999.

\bibitem[Chevaleyre et~al.(2006)Chevaleyre, Dunne, Endriss, Lang, Lemaitre,
  Maudet, Padget, Phelps, Rodriguez-Aguilar, and Sousa]{chevaleyre2006issues}
Yann Chevaleyre, Paul~E Dunne, Ulle Endriss, J{\'e}r{\^o}me Lang, Michel
  Lemaitre, Nicolas Maudet, Julian Padget, Steven Phelps, Juan~A
  Rodriguez-Aguilar, and Paulo Sousa.
\newblock Issues in multiagent resource allocation.
\newblock \emph{Informatica}, pages 3--31, 2006.

\bibitem[De~Jong et~al.(2008)De~Jong, Uyttendaele, and Tuyls]{de2008learning}
Steven De~Jong, Simon Uyttendaele, and Karl Tuyls.
\newblock Learning to reach agreement in a continuous ultimatum game.
\newblock \emph{J. Artif. Intell. Res}, 33:\penalty0 551--574, 2008.

\bibitem[Deng et~al.(2014)Deng, Liu, Sadiq, and Deng]{deng2014impact}
Xinyang Deng, Qi~Liu, Rehan Sadiq, and Yong Deng.
\newblock Impact of roles assignation on heterogeneous populations in
  evolutionary dictator game.
\newblock \emph{Sci. Rep.}, 4:\penalty0 6937, 2014.

\bibitem[Dorogotsev et~al.(2001)Dorogotsev, Mendes, and
  Samukhin]{Dorogotsev2001}
S.~N. Dorogotsev, J.~F.~F. Mendes, and A.~N. Samukhin.
\newblock Size-dependent degree distribution of a scale-free growing network.
\newblock \emph{Phys Rev E}, 63, 2001.

\bibitem[Fehr and Fischbacher(2003)]{fehr2003nature}
Ernst Fehr and Urs Fischbacher.
\newblock The nature of human altruism.
\newblock \emph{Nature}, 425\penalty0 (6960):\penalty0 785, 2003.

\bibitem[Fehr and Schmidt(1999)]{fehr1999theory}
Ernst Fehr and Klaus~M Schmidt.
\newblock A theory of fairness, competition, and cooperation.
\newblock \emph{Q. J. Econ}, 114\penalty0 (3):\penalty0 817--868, 1999.

\bibitem[Gini(1921)]{gini1921measurement}
Corrado Gini.
\newblock Measurement of inequality of incomes.
\newblock \emph{The economic journal}, 31\penalty0 (121):\penalty0 124--126,
  1921.

\bibitem[Grimm et~al.(2017)Grimm, Feicht, Rau, and Stephan]{grimm2015impact}
Veronika Grimm, Robert Feicht, Holger Rau, and Gesine Stephan.
\newblock On the impact of quotas and decision rules in ultimatum collective
  bargaining.
\newblock \emph{Eur. Econ. Rev.}, 100:\penalty0 175--192, 2017.

\bibitem[G{\"u}th et~al.(1982)G{\"u}th, Schmittberger, and
  Schwarze]{guth1982experimental}
Werner G{\"u}th, Rolf Schmittberger, and Bernd Schwarze.
\newblock An experimental analysis of ultimatum bargaining.
\newblock \emph{J. Econ. Behav. Organ.}, 3\penalty0 (4):\penalty0 367--388,
  1982.

\bibitem[Jennings et~al.(2001)Jennings, Faratin, Lomuscio, Parsons, Sierra, and
  Wooldridge]{jennings2001automated}
Nicholas~R Jennings, Peyman Faratin, Alessio~R Lomuscio, Simon Parsons, Carles
  Sierra, and Michael Wooldridge.
\newblock Automated negotiation: prospects, methods and challenges.
\newblock \emph{Group Decis Negot}, 10\penalty0 (2):\penalty0 199--215, 2001.

\bibitem[Lorenz(1905)]{lorenz1905methods}
Max~O Lorenz.
\newblock Methods of measuring the concentration of wealth.
\newblock \emph{Publications of the American statistical association},
  9\penalty0 (70):\penalty0 209--219, 1905.

\bibitem[Lynch et~al.(2018)Lynch, Tran-Thanh, Santos,
  et~al.]{lynch2018fostering}
Simon Lynch, Long Tran-Thanh, Francisco~C Santos, et~al.
\newblock Fostering cooperation in structured populations through local and
  global interference strategies.
\newblock In \emph{Proc of IJCAI'18}, 2018.

\bibitem[Nowak et~al.(2000)Nowak, Page, and Sigmund]{nowak2000fairness}
Martin~A Nowak, Karen~M Page, and Karl Sigmund.
\newblock Fairness versus reason in the ultimatum game.
\newblock \emph{Science}, 289\penalty0 (5485):\penalty0 1773--1775, 2000.

\bibitem[Page et~al.(2000)Page, Nowak, and Sigmund]{page2000spatial}
Karen~M Page, Martin~A Nowak, and Karl Sigmund.
\newblock The spatial ultimatum game.
\newblock \emph{Proc R Soc B}, 267\penalty0 (1458):\penalty0 2177--2182, 2000.

\bibitem[Pinheiro and Santos(2018)]{pinheiro2018local}
Fl{\'a}vio~L Pinheiro and Fernando~P Santos.
\newblock Local wealth redistribution promotes cooperation in multiagent
  systems.
\newblock In \emph{Proc of AAMAS'18}, pages 786--794, 2018.

\bibitem[Pritchett and Genton(2017)]{pritchett2017negotiated}
Amy~R Pritchett and Antoine Genton.
\newblock Negotiated decentralized aircraft conflict resolution.
\newblock \emph{IEEE T Intell Transp}, 19\penalty0 (1):\penalty0 81--91, 2017.

\bibitem[Raghunandan and Subramanian(2012)]{raghunandan2012sustaining}
MA~Raghunandan and CA~Subramanian.
\newblock Sustaining cooperation on networks: an analytical study based on
  evolutionary game theory.
\newblock In \emph{Proc of AAMAS'12}, pages 913--920, 2012.

\bibitem[Rahwan(2019)]{rahwan2019machine}
Iyad et~al. Rahwan.
\newblock Machine behaviour.
\newblock \emph{Nature}, 568\penalty0 (7753):\penalty0 477--486, 2019.

\bibitem[Rand et~al.(2013)Rand, Tarnita, Ohtsuki, and Nowak]{rand2013evolution}
David~G Rand, Corina~E Tarnita, Hisashi Ohtsuki, and Martin~A Nowak.
\newblock Evolution of fairness in the one-shot anonymous ultimatum game.
\newblock \emph{Proc. Natl. Acad. Sci. USA}, 110\penalty0 (7):\penalty0
  2581--2586, 2013.

\bibitem[Ranjbar-Sahraei et~al.(2014)Ranjbar-Sahraei, Bou~Ammar, Bloembergen,
  Tuyls, and Weiss]{ranjbar2014evolution}
Bijan Ranjbar-Sahraei, Haitham Bou~Ammar, Daan Bloembergen, Karl Tuyls, and
  Gerhard Weiss.
\newblock Evolution of cooperation in arbitrary complex networks.
\newblock In \emph{AAMAS'14}, pages 677--684, 2014.

\bibitem[Salazar et~al.(2011)Salazar, Rodriguez-Aguilar, Arcos, Peleteiro, and
  Burguillo-Rial]{salazar2011emerging}
Norman Salazar, Juan~A Rodriguez-Aguilar, Josep~Ll Arcos, Ana Peleteiro, and
  Juan~C Burguillo-Rial.
\newblock Emerging cooperation on complex networks.
\newblock In \emph{Proc of AAMAS'11}, pages 669--676, 2011.

\bibitem[Santos et~al.(2015)Santos, Santos, Paiva, and
  Pacheco]{santos2015evolutionary}
Fernando~P Santos, Francisco~C Santos, Ana Paiva, and Jorge~M Pacheco.
\newblock Evolutionary dynamics of group fairness.
\newblock \emph{J Theor Biol}, 378:\penalty0 96--102, 2015.

\bibitem[Santos et~al.(2016)Santos, Santos, Melo, Paiva, and
  Pacheco]{santos2016dynamics}
Fernando~P Santos, Francisco~C Santos, Francisco~S Melo, Ana Paiva, and Jorge~M
  Pacheco.
\newblock Dynamics of fairness in groups of autonomous learning agents.
\newblock In \emph{AAMAS'16 Workshops, Best Papers}, pages 107--126. Springer,
  2016.

\bibitem[Santos et~al.(2017)Santos, Pacheco, Paiva, and
  Santos]{santos2017structural}
Fernando~P Santos, Jorge~M Pacheco, Ana Paiva, and Francisco~C Santos.
\newblock Structural power and the evolution of collective fairness in social
  networks.
\newblock \emph{PloS ONE}, 12\penalty0 (4):\penalty0 e0175687, 2017.

\bibitem[Santos et~al.(2019)Santos, Pacheco, Paiva, and
  Santos]{santos2019evolution}
Fernando~P Santos, Jorge~M Pacheco, Ana Paiva, and Francisco~C Santos.
\newblock Evolution of collective fairness in hybrid populations of humans and
  agents.
\newblock In \emph{Proc of AAAI'19}, volume~33, pages 6146--6153, 2019.

\bibitem[Santos et~al.(2008)Santos, Santos, and Pacheco]{santos2008social}
Francisco~C Santos, Marta~D Santos, and Jorge~M Pacheco.
\newblock Social diversity promotes the emergence of cooperation in public
  goods games.
\newblock \emph{Nature}, 454\penalty0 (7201):\penalty0 213, 2008.

\bibitem[Shirado and Christakis(2017)]{shirado2017locally}
Hirokazu Shirado and Nicholas~A Christakis.
\newblock Locally noisy autonomous agents improve global human coordination in
  network experiments.
\newblock \emph{Nature}, 545\penalty0 (7654):\penalty0 370--374, 2017.

\bibitem[Takesue et~al.(2017)Takesue, Ozawa, and
  Morikawa]{takesue2017evolution}
Hirofumi Takesue, Akira Ozawa, and So~Morikawa.
\newblock Evolution of favoritism and group fairness in a co-evolving
  three-person ultimatum game.
\newblock \emph{Europhysics Letters (EPL)}, 118\penalty0 (4):\penalty0 48002,
  2017.

\bibitem[Traulsen et~al.(2006)Traulsen, Nowak, and
  Pacheco]{traulsen2006stochastic}
Arne Traulsen, Martin~A Nowak, and Jorge~M Pacheco.
\newblock Stochastic dynamics of invasion and fixation.
\newblock \emph{Phys Rev E}, 74\penalty0 (1):\penalty0 011909, 2006.

\bibitem[Tuyls and Parsons(2007)]{tuyls2007evolutionary}
Karl Tuyls and Simon Parsons.
\newblock What evolutionary game theory tells us about multiagent learning.
\newblock \emph{Artif. Intell.}, 171\penalty0 (7):\penalty0 406--416, 2007.

\bibitem[Weibull(1997)]{weibull1997evolutionary}
J{\"o}rgen~W Weibull.
\newblock \emph{Evolutionary game theory}.
\newblock MIT press, 1997.

\bibitem[Wu et~al.(2013)Wu, Fu, Zhang, and Wang]{wu2013adaptive}
Te~Wu, Feng Fu, Yanling Zhang, and Long Wang.
\newblock Adaptive role switching promotes fairness in networked ultimatum
  game.
\newblock \emph{Sci. Rep.}, 3:\penalty0 1550, 2013.

\bibitem[Zisis et~al.(2015)Zisis, Di~Guida, Han, Kirchsteiger, and
  Lenaerts]{zisis2015generosity}
Ioannis Zisis, Sibilla Di~Guida, TA~Han, Georg Kirchsteiger, and Tom Lenaerts.
\newblock Generosity motivated by acceptance-evolutionary analysis of an
  anticipation game.
\newblock \emph{Sci. Rep.}, 5:\penalty0 18076, 2015.

\end{thebibliography}

%%%%%%%%%%%%%%%%%%%%%%%%%%%%%%%%%%%%%%%%%%%%%%%%%%%%%%%%%%%%%%%%%%%%%%%%

\end{document}